\documentclass[12pt]{iopart}
\usepackage{graphicx}
\usepackage[usenames,dvipsnames]{color}
\usepackage[a4paper=true,ps2pdf=true]{hyperref}
\usepackage{sidecap}

\hypersetup{
    colorlinks = true,
    linkcolor = blue,
    anchorcolor = green,
    citecolor = blue,
    filecolor = red,
    urlcolor = red
}

\begin{document}

\title[Lattice Jahn-Teller model]{Quantum phase crossover and  chaos in
generalized Jahn-Teller lattice model}

\author{Eva Majern\'{\i}kov\'a$^{1}$ and Serge Shpyrko$^{2}$\footnote{email: serge\_shp@yahoo.com}}

\address
{${}^{1}$ Institute of Physics, Slovak Academy of Sciences,
D\'ubravsk\'a cesta 9, SK-84 511 Bratislava, Slovak Republic}

\address
{${^{2}}$Institute for Nuclear Research, Ukrainian Academy of
Sciences, pr. Nauki 47, Kiev, Ukraine}

\ead{eva.majernikova@savba.sk}


 \begin{abstract}
The generalized multispin Jahn-Teller  model on a finite lattice or
formally equivalent Dicke model extended to two long-wavelength
coherent bosons of different frequencies is shown to exhibit a
crossover between the polaron-modified "quasi-normal" and the
squeezed "radiation"  domains. We investigate effects of two kinds
of interfering fluctuations on the phase crossover and on
statistical characteristics of boson complex spectra:
 (i) Fluctuations in the {\it electron} subsystem- finite-size quantum fluctuations-
  are responsible for the dephasing of the
coherence in the radiation domain and for the moderate occupation of
the excited states in the normal domain. In the quasiclassical
limit, radiation phase implies  existence of a coherent acoustic
super-radiant phase. (ii) Level-spacing fluctuations in excited {\it
boson} level subsystem  with strong level repulsions. Related
probability distributions are shown to be non-universally spread
between the limiting universal Wigner-Dyson  and  Poisson
distributions.  We proved that the difference in  boson frequencies
is responsible for reaching the most stochastic limit of the
Wigner-Dyson distribution. Instanton lattice as a sequence of
tunneling events in the most chaotic radiation domain exhibits
maximal number of level-avoidings (repulsions). The non-universality
of the distributions is caused by boson correlations which compete
the level repulsions.
\end{abstract}


\submitto{\JPA}

 \maketitle

\section{Introduction}
\label{intro} The class of spin-boson lattice models
 with one \cite{Graham:1984:a,Graham:1984:b,Lew:1991,
 Cibils:1995,Dicke,Wang:1973,Littlewood:2000,Emary:2003:a,Emary:2003:b}
 or two boson modes of different local symmetry
\cite{Haake,Tolkunov:2007,Pfeifer,Larson:2008} gained a long-term
interest in condensed matter physics and quantum optics. They
exhibit a rich variety of interesting statistical properties as
spectral chaos and quantum and thermodynamic phase transitions. The
complex (chaos bearing) excited energy spectra are a consequence of
multiple level repulsions and avoidings  \cite{Guhr}. Recently,
discovery of the exciton-polariton Bose-Einstein condensation in
semiconductor microcavities
\cite{Littlewood:2000,Kasprzak:2006,Eastham:2006,Keeling:2007}
contributed to this fascinating field.

 An array of two-level atoms localized on a lattice and
 dipole-interacting with one long-wavelength boson (photon) mode --
 the Dicke model \cite{Dicke} -- has been found to
exhibit quantum and thermodynamic spontaneous phase transition in adiabatic approximation
\cite{Dicke,Wang:1973,Littlewood:2000,Emary:2003:a,Emary:2003:b,Keeling:2007,Love:2008}
from the effectively unexcited "normal" phase to the "super-radiant"
phase, a macroscopically excited and highly collective state. A
collective spontaneous emission of coherent radiation is due to the
cooperative interaction of a large number of two-level localized
atoms all excited to the upper state. Moreover, the Dicke model
 has been found to exhibit the Wigner-Dyson
probability distribution of the nearest neighbour level spacings
(NNLS) of excited boson spectra \cite{Emary:2003:b}. Generally, this
type of fluctuations was intensively studied by the quantum chaology
\cite{Guhr} for understanding the complex excited spectra of
many-body systems \cite{Dyson}.

There are several motivations for the present generalization to the
multispin  JT model and different
frequencies of two boson modes:

(a) Our preceding investigations \cite{Majernikova:2002,
Majernikova:2003}  of the ground state of the Jahn-Teller (JT) model
resulted in finding features analogous to those of the Dicke model;

(b) Investigation of characteristics of quantum chaos in
E$\otimes(b_1+b_2)$ JT spectra \cite{Majernikova:2008} left open
questions: we expect further enhancement of the
 level-spacing fluctuations due to the difference in the mode
frequencies  and related impact on characteristics of chaos;

(c) Potential applications for another two-level  finite-lattice
class of models with localization-delocalization phase transition,
e.g., in (i) ultracold Bose gas in optical lattices: The
Bose-Einstein condensation exhibits the phase coherence of Bose
atoms destroyed by the suppression of the tunneling due to the
localization by the strong lattice potential
 \cite{Ziegler:2005,Roth:2004};
 (ii) recently discovered Bose-Einstein condensation
of exciton polaritons (bosons), coupled light-matter bosonic quasiparticles
\cite{Kasprzak:2006} in semiconductor microcavities. Theoretical
investigation of the phase transition to the polariton condensation
state, including finite-size fluctuations and quantum effects vs
finite temperatures were reported
\cite{Littlewood:2000,Eastham:2006,Love:2008}.
The finite-size effects for the polariton condensation phase
transition within the Dicke model were studied by Eastham et al
\cite{Eastham:2006} and  scaling corrections to the critical
exponents for the Dicke model at the crossover  were studied by
Vidal et al \cite{Vidal:2006}.

The finite-size aspect appears naturally in our system through
 two kinds of fluctuations: \\
 1.  Finite-size fluctuations in {\it electron} subsystem due to the
finiteness of the lattice and\\
2. fluctuations in {\it boson } subsystem -- level spacing
fluctuations in the discrete set of excited levels of the boson
spectra. The signs of complexity are the multiple quantum level
repulsions and respective level avoidings. They interplay with
quantum correlations and are responsible for the spectral
characteristics with the signatures of quantum chaos. The quantum
correlations competing the level repulsions include (i) mode
selfinteraction (squeezing) resulting in reduction of the mode
frequency and stabilizing the ground state. (ii) Quantum
correlations between two different modes (entanglement). The
two-boson (squeezed) coherent states are analogous to the
squeezed  coherent states in quantum optics introduced by
 Yuen \cite{Yuen}.

(The fluctuations due to the discreteness of spectra even without
the signs of complexity are known to be present in a number of
coupled quantum electron-phonon systems that do not admit the use of
the adiabatic approximation \cite{Lowen:1988}. They cause smooth
crossover in localization-delocalization phase transition not
accompanied by a non-analytical change of the ground state).

In Section II we define the model under investigation  and  review
the representation of the Dicke multispin collective states.
 In Section III  we investigate numerically and
analytically the ground state. For analytical calculations we use
the Holstein-Primakoff bosonisation Ansatz for the multispin
operators.
 In Section IV we present an approximate analytical
investigation of the role of nonlinear terms
accompanying the finite-size effects. It provides a useful, though
approximate, insight into a scenario of interaction of three
effective oscillators especially on a formation of the mixed phase.

 In Section V, after  a brief summary of realms of the theory of
quantum spectral chaos, we show  numerical results on statistical
signatures of the quantum chaos - the
probability distributions of nearest neighbour level spacings (NNLS),
spectral entropies and spectral densities
for different numbers of lattice sites, frequencies and coupling
parameters. The interference of the finite-size fluctuations in
electron space with boson level spacing fluctuations will be
demonstrated by the statistical characteristics implying the quantum
chaotic behaviour.
 A qualitative difference between the present model and the one-boson Dicke model
 will be illustrated by comparison of the nearest NNLS distributions of both models.

  Considering possible applications, in Conclusion  we propose several
interpretations of experimental results.  Generally, the concepts of
the  model above can be applied for nanostructures where the quantum
effects are found rather detectable \cite{Eisert:2004}.

\section{Model}
\label{model}

We consider a finite lattice array of localized JT molecules - two
electron levels at each lattice site dipole- coupled to two vibron
modes of different symmetry against reflection  with different
coupling strengths and different frequencies. The simplest
(one-boson) E$\otimes \beta$ version of the JT model, in the
collective coordinate approximation, reveals a formal analogy to the
one-boson Dicke model \cite{Dicke} or to the exciton or dimer model.

The JT Hamiltonian generalized to multiple dimensionality of the
(pseudo)spin space and different boson frequencies is of the form
\begin{equation}
H = (\Omega_1 a_{1}^{\dag}a_{1}  + \Omega_2 a_{2}^{\dag}a_{2}) I +
\frac{1}{\sqrt {N}}\sum\limits_{i=1}^N  \Big[\alpha (a_{1
i}^{\dag}+a_{1i})2 s_z^{(i)}+\beta (a_{2 i}^{\dag}+a_{2i
})(s_{+}^{(i)}+s_{-}^{(i)})\Big].  \label{h}
\end{equation}

The wave functions of the localized JT molecules  do not overlap.
The bosons represent intramolecular vibrons:  the antisymmetric
against reflection mode $\Omega_1, \ a_1$ assists the splitting  of
the level (in the Dicke model the level separation is a model
parameter). The symmetric boson mode $a_2$ assists the tunneling
between the levels. (Evidently, the simplest  one boson version of
the JT model (E$\otimes \beta$) and the Dicke  model are related by
a unitary transformation - rotation in the pseudospin $2\times 2$
space). The dipole electron-boson coupling causes the
non-conservation of the  coherent bosons number and, consequently,
implies the nonintegrability of Hamiltonian (\ref{h}) as opposed
to the Jaynes-Cummings model.

The on-site displacements are homogeneous along the lattice: The
sites (levels) are equally displaced within the $N=2j$ multiplicity,
so that we can omit the index $i$ of the dipole momentum operator
 in (\ref{h}). Then, using the Dicke
collective ($N$-dimensional) spin variables $J_z=\sum\limits_{i}
s^{(i)}_z$, $J_+=\sum\limits_i s^{(i)}_+$, $J_-=\sum\limits_{i}
s^{(i)}_-$, Hamiltonian (\ref{h}) can be rewritten to the Dicke-like
form

\begin {equation}
H   = \left(\Omega_1 a_{1}^{\dag}a_{1}  +  \Omega_2 a_{2}^{\dag}a_{2}\right) I
 + \frac{\alpha}{\sqrt{2j}}
\left( a_{1}^{\dag}+a_{1}\right)\cdot 2J_{z}
 +\frac{\beta}{\sqrt{2j}} \left(a_2^{\dag}+a_2\right)\left( J_+ + J_- \right)\,. \label{h1}
\end{equation}

 The operators $J_{\pm}$, $J_z$ satisfy the same commutation relations as the individual spins:
\begin{equation}
 [J_z, J_{\pm}]=\pm J_{\pm}\,; \quad [J_+,J_-]=2 J_z
\label{spin:alg}
\end{equation}
where $J_x={\displaystyle \frac{1}{2}}(J_+ +J_-), J_y=-{\displaystyle \frac{i}{2}}(J_+-J_-)$.

  The scaling by $1/\sqrt
{2j}$ in (\ref{h1}) accounts for lengths of the dipoles
$2j=1,2,\dots N$. It was chosen equal for both interaction terms for
symmetry, but eventual rescaling of one of the coupling constants
$\alpha, \ \beta$ is evidently possible, if necessary.

 The model (\ref{h1}) can be also considered as a
generalization of one-boson Dicke model
\begin{equation}
H_D=\Omega  a^{\dag}a  I
 + \omega_0 J_{z}
 +\frac{\lambda}{\sqrt{2j}} (a^{\dag}+a)\left( J_+ + J_- \right)\,. \label{HD}
\end{equation}

\subsection{Representation of the Dicke states }
The  Dicke states are represented by the spin operators of higher
dimensions \cite{Dicke}. Each Dicke state is labeled as
$|j,m\rangle$ with the discrete index $m$ ranging (at fixed $j$)
from $-j,-j+1,\dots$ to $j$. The spin operators in the Dicke space
are defined as follows:
\begin{eqnarray}
 J_z |j,m\rangle = m |j,m\rangle\,;\   J_{\pm}|j,m\rangle= \sqrt{j(j+1)-m(m\pm 1)}
 |j,m\pm 1\rangle\,;\nonumber\\   J^2 |j,m\rangle=j(j+1) |j,m\rangle\,.
\label{Dicke:alg}
\end{eqnarray}

The subspaces  $j$ are independent and, hence, one can consider them
separately keeping $j$ fixed.
 For the chain containing $N$ atoms possible values of $j$ range as $0,\, 1,\dots , N/2$ for
 $N$ even and $1/2,\, 3/2, \dots , N/2$ for  $N$ odd; for given $j$ the operators $J_i$ are matrices
 with dimensions $2j+1$ (representations of the $SU(2)$ group). In what follows we shall
always take into consideration only the subspace with largest
$j=N/2$ for each number of atoms in a chain.

The  operator of parity $ \Pi = \exp \{i\pi (a_i^{\dag}a_i +
J_z+j)\},\ i=1,2$ commutes with Hamiltonian (\ref{HD}).
 For the JT molecule  with
only two electron levels  ($j=1/2$) and equal boson frequencies this
fact reflects itself in the conserved parity $p=\pm 1$
\cite{Long:1958,Majernikova:2003} as an additional good quantum
number.  For parities $+1$ and $-1$ the spectra of that model are
identical. As we shall see, for the present generalization this is
the case of even number of molecules, that is for the main Dicke
numbers $j=1/2,3/2,\dots$.

\section{Ground state}
\label{ground}

The method of collective pseudospin variables  provides a useful
insight into the scenario of the interplay of interactions in the
system of  three effective oscillators.
 The Holstein-Primakoff Ansatz \cite{Primakoff}   for the
collective pseudospin operators (\ref{spin:alg}) reads

 \begin{equation}
  J_z= b^{\dag}b-j,\ J_+=
b^{\dag}\sqrt{2j-b^{\dag}b}, \ J_-= \sqrt{2j-b^{\dag}b}\,b .
\label{hp}
 \end{equation}

 Here $b$ are fictitious
  boson operators which satisfy boson commutation rules
$[b,b^\dag]=1$; this representation of the
spin algebra preserves exactly the commutation relations
(\ref{spin:alg}) and makes it possible to convert the system to two
or three coupled quantum oscillators as will be shown below. (Let us
remark that the Holstein-Primakoff Ansatz (\ref{hp}) breaks the
symmetry of Hamiltonian (\ref{h1}) with respect to a simultaneous
exchange of $\alpha\leftrightarrow \beta$, $\Omega_1\leftrightarrow
\Omega_2$ and $J_z\leftrightarrow J_x=J_++J_-$ well known for the
Jahn-Teller model with $2j=1$.)

Linear terms in Hamiltonian (\ref{h1}) can be excluded by use of the
coherent state representation with boson displacements chosen
variationally providing $\langle b^{\dag}b\rangle/2j \ll 1$, i.e.
close to saturation \cite{Primakoff}.
Let us displace the operators involved as follows
\begin{equation}
 a_1^{\dag}=
c^{\dag}_1+\sqrt{\alpha_1}, \quad a_2^{\dag}=  c^{\dag}_2+
\sqrt{\alpha_2},   \quad b^{\dag}=
d^{\dag}-\sqrt{\delta}\,.\label{shift}\end{equation}

Applying the Holstein-Primakoff Ansatz (\ref{hp}), setting
(\ref{shift}) into  (\ref{h1}) and  expanding the square root
expressions in (\ref{hp})  yields the form
\begin{eqnarray}
H\approx \Omega_1\left( c_1^{\dag}c_1+1/2 \right) + \Omega_2\left(
c_2^{\dag}c_2+1/2 \right) + \Omega_1\alpha_1+\Omega_2\alpha_2
+\Omega_1
\sqrt{\alpha_1}\left(c_1^{\dag}+c_1\right)\nonumber\\+\Omega_2
\sqrt{\alpha_2}\left( c_2^{\dag}+ c_2\right)
 +\frac{2\alpha}{\sqrt {2j}}\left[(c_1^{\dag}+c_1)d^{\dag}d
+2\sqrt{\alpha_1}d^{\dag}d
-\sqrt{\delta}(c_1^{\dag}+c_1)(d^{\dag}+d) \right.\nonumber\\
\left.-2\sqrt{\alpha_1\delta}(d^{\dag}+d)+
(\delta-j)(c_1^{\dag}+c_1)
 +2\sqrt{\alpha_1}(\delta-j)\right ]\nonumber\\+ \frac{ \beta}{\sqrt{2j}}(
c_2^{\dag}+ c_2+2\sqrt{ \alpha_2}) \cdot k \left \{-2\sqrt{\delta} +
(1-\delta/k^2)(d^{\dag}+d)+\frac{\sqrt{\delta}}{k^2}d^{\dag}d\right.
\nonumber\\\left.+\frac{1}{2k^2}\left[-(d^{\dag
2}d+d^{\dag}d^2)+\sqrt{\delta}((d^{\dag}+d)^2-1)
 \right ]\right\},\quad k\equiv\sqrt{2j-\delta}\,. \ \
\label{H}
\end{eqnarray}

From the condition of elimination of the terms linear in boson
operators from (\ref{H}) three identities for the parameters
$\alpha_i$ and $\delta$ follow ($k\neq 0$):
 \begin{eqnarray}
&\Omega_1\sqrt{\alpha_1}= -\frac{ 2\alpha}{\sqrt{2 j} }\,(\delta-j)\,, \nonumber \\
&\Omega_2 \sqrt{\alpha_2}= \frac{\beta\sqrt{\delta}}{\sqrt{2j}}\,
\frac{4j-2\delta+1/2}{\sqrt{2j-\delta }}\,, \label {displ} \\
&\sqrt{\delta}(\delta-j)\left
(4\Omega_2\,\alpha^2-2\Omega_1\,\beta^2\,\frac{4j-2\delta+1/2}{2j-\delta}\right
)=0 \,. \nonumber
\end{eqnarray}

For finite $j$, this set of equations implies three solutions
provided $ \Omega_2\,\alpha^2\neq \Omega_1\, \beta^2(1+1/8j)$ as
follows: \numparts
\begin{eqnarray}  \sqrt{\alpha_1}=\frac{\alpha}{\Omega_1}\sqrt {2j}, \ \alpha_2=0 , \,
\delta=0\, \label{sol1}\\
     \alpha_1= 0, \  \sqrt
{\alpha_2}=\frac{\beta}{\Omega_2}\sqrt{2 j}\left(1+\frac{1}{4j}\right) , \  \delta=j  ,  \label{sol2} \\
 \sqrt{\alpha_1}= -\frac{\alpha}{\Omega_1}\sqrt{2j}
(1-2\bar\mu)\,, \ \sqrt{\alpha_2}=\frac{2\alpha^2}{\Omega_1 \beta}
\sqrt{2j} \sqrt{\bar{\mu}(1-\bar{\mu})}, \ \delta=2j(1-\bar\mu
)\,, \qquad \label{sol3}
\end{eqnarray}
\endnumparts
 where in (\ref{sol3})  $\bar\mu =
{\displaystyle\frac{\bar{\beta}^2}{8j\left(\alpha^2-{\bar{\beta}}^2\right)}}
 <1$,  and
$\alpha > \bar{\beta} \equiv \sqrt{\Omega_1/\Omega_2}\,\beta$.

The first two solutions are similar to that reported for the Dicke
model of superradiance
 \cite{Emary:2003:a,Emary:2003:b,Tolkunov:2007,Brandes:2005} in the limit
 $j\rightarrow \infty$, so that in this
limit they account for the common normal and super-radiant phases.
In this case the condition
$\alpha\sqrt{\Omega_2}=\beta\sqrt{\Omega_1}$ would represent the
point of the quantum phase transition. The phase transition for the
present model  means transition from the state (\ref{sol1}) with
zero macroscopic occupation to the fully excited $q$-bit state
(\ref{sol2}). In the semiclassical limit this would imply possible
concept of a "superradiant phonon state" (\ref{sol2}).

On the other hand, for finite $j$, third solution (\ref{sol3})
represents specific finite-size effects; the thermodynamic limit
(and thus the phase transition) is forbidden due to the condition
$0<\bar\mu<1$. Instead of
 the phase transition between the normal (\ref{sol1}) and radiation (\ref{sol2})
 phase, there appears   a crossover (intermediate) domain within the "normal" domain
 (\ref{sol3}). In this case all three oscillators  are coupled: the parameter $\bar\mu$ (\ref{sol3})
 is a measure of the coupling. Evidently, in the limit $j\rightarrow\infty$ the phase (\ref{sol3}) vanishes.

\subsection{Case 1: Polaron self-trapping in the normal domain}
\label{ground:1}

 In this case (\ref{sol1}) we expect an impact of the additional oscillator
 $a_1$ (here its fluctuations $c_1$ (\ref{shift})) on the behaviour of the normal phase.

 The solution  (\ref{sol1}) represents  the normal phase with average zero number of
 macroscopically excited electron level bosons $\delta=0$.
  Hamiltonian (\ref{H}) related to this solution reads
\begin{eqnarray} H_1= \Omega_1\left(c_1^{\dag}c_1+1/2
\right) + \Omega_2\left( c_2^{\dag}c_2+1/2 \right)
+\frac{4\alpha^2}{\Omega_1}(d^{\dag}d-j/2)+ \beta (
{c_2}^{\dag}+  c_2)(d^{\dag}+d) \nonumber\\
+\frac{2\alpha}{\sqrt{2j}}(c_1^{\dag}+c_1)d^{\dag}d -\frac{
\beta}{4j }( c_2^{\dag}+c_2)(d^{\dag 2} d +d^{\dag } d^2)\,.\qquad
\label{H1}
\end{eqnarray}

\begin{SCfigure}[][htb]
\includegraphics[width=0.6\hsize]{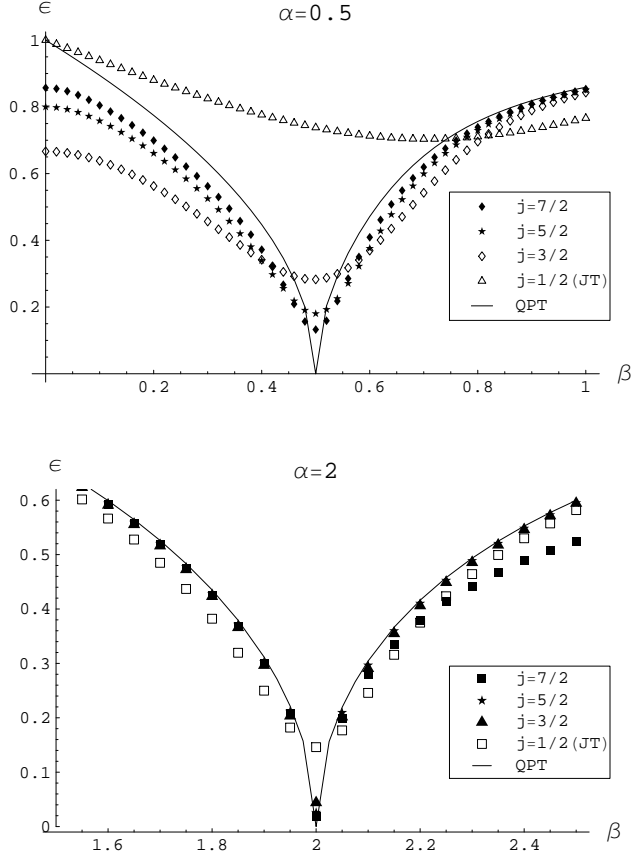} \hspace{-3em}
\caption{ Crossover between the normal and radiation domain in the
generalized  JT lattice model for $\alpha=0.5$, $2$ and resonance
case $\Omega_1=\Omega_2=1$. The numerical results for the excitation
energy of the first excited state for different $j$ are shown, the
solid line in each figure is the analytical result for asymptotic
QPT (\ref{eps1}), (\ref{eps2}) valid for $j\to\infty$. The cusp-like
behavior close to the  point of the crossover
$\omega_1\equiv4\alpha^2/\Omega_1=4\beta^2/\Omega_2^2\equiv\omega_2$
appears already for relatively small number of sites, e.g.,  for
$j=7/2$. For small $j$, the fluctuations smooth the cusp especially
at weak couplings. The non-symmetry about the critical point by the
reduction of the energy in the radiation domain for $\alpha=0.5$ is
due to the squeezing. For the case of different
$\Omega_1\neq\Omega_2$ the picture is qualitatively the same, there
occurs a shift of the transition inwards the phase with higher
$\Omega_i$.} \label{boson-fig1}
\end{SCfigure}

Here,  the level oscillator  acquires the frequency of a polaron-
dressed oscillator $\omega_1=4\alpha^2/\Omega_1$ which results from
the self-trapping by the additional mode in contrast with respective
factor of the Dicke model  $\omega_0$.
 Explicit interaction of the
oscillator modes with the level bosons contribute only to excited
states  $\sim 1/\sqrt j $ in (\ref{H1}).  For large $j$ only linear
terms persist (first line in (\ref{H1})) and Hamiltonian (\ref{H1})
 can be easily diagonalized by a rotation in the plane of coupled
operators $c_2$ and $d$ \cite{Emary:2003:b} to yield the form
\begin{equation}
 H_{1d}= \Omega_1 c_1^{\dag}c_1+ \epsilon_{1,
+}C_2^{\dag}C_2 +  \epsilon_{1,
-}C_3^{\dag}C_3-2\alpha^2j/\Omega_1\, \label{E1}
\end{equation}
with new effective oscillators $C_2$, $C_3$. The excitation energies
of the system $\epsilon_{1,\pm}$ in the thermodynamic limit are

\begin{equation}
\epsilon_{1, \pm}^2= \frac{1}{2}\Big(\Omega_2^2 +\omega_1^2\pm
\left[(\Omega_2^2-\omega_1^2)^2+ 64\alpha^2 \beta^2\Omega_2/\Omega_1
\right]^{1/2}\Big)\,.\label{eps1}
\end{equation}
 From (\ref{eps1}), the solution for $\epsilon_1 $ exists
provided $\omega_1> 4\beta^2/\Omega_2 \equiv \omega_2$, or if
$\alpha\sqrt{\Omega_2} > \beta\sqrt{\Omega_1}$. This
 phase is identified as the normal phase of the Dicke-like
model without macroscopic excitations, Figure \ref{boson-fig1}. The
effect of different initial frequencies $\Omega_1\neq \Omega_2$ is
just the shift of the phase transition point supporting the phase
with the smaller of $\Omega_i$.

From (\ref{H1}),  critical Hamiltonian $H_1^{crit}= H_1
(\beta_c=\alpha\sqrt{\Omega_2/\Omega_1})$ at the point of the phase
transition yields
\begin{equation}
H_1^{crit}=\sqrt{\Omega_2^2+ \left(4\alpha^2/\Omega_1\right)^2}
\left(C^{\dag}_2 C_2+1/2\right) + \Omega_1
\left(c_1^{\dag}c_1+1/2\right) -2\alpha^2 j/\Omega_1\,.
\label{H1crit}
\end{equation}

  The coupled undisplaced oscillators $a_2$ and $d$ form an
effective single oscillator of the frequency $\sqrt{\Omega_2^2+\left
(4\alpha^2/\Omega_1\right )^2}$. The mixing results from the
coupling due to the assisting oscillator $a_2$ between the levels
split by the oscillator $a_1$.

\subsection{Case 2: Squeezing in the radiation domain.}
\label{ground:2}

The second solution (\ref{sol2}) identified further with the
radiation domain, can be treated on the same footing. For this
solution the level bosons (\ref{hp}) are displaced by $\delta=j$
which represents the number of excited states within the domain.
 Respective  Hamiltonian (\ref{H}) yields
 \begin{equation}\label{H2}
   H_2=H_{2}^{(1)}+ H_{2}^{(2)},
 \end{equation}
where
\begin{eqnarray}\label{H21}
H_{2}^{(1)} = \Omega_1\left( c_1^{\dag}c_1+1/2 \right) +
\Omega_2\left( c_2^{\dag}c_2 +1/2\right)
+4\beta^2\left(1+1/4j\right) d^{\dag}d /\Omega_2\nonumber
\\-
  \sqrt{2}\alpha (c_1^{\dag}+c_1)(d^{\dag}+d) +
\beta^2\left(1+1/4j\right)(d^{\dag 2}+d^2)/\Omega_2 \ ,\qquad \
\end{eqnarray}
and a fluctuation part of nonlinear interaction terms  $O( 1/\sqrt
j),\ O(1/j)$,
\begin{eqnarray}
 H_{2}^{(2)}=   \frac{1}{\sqrt{2j}}\left(2\alpha(c_1^{\dag}+c_1)d^{\dag}d +
\beta(c_2^{\dag}+c_2)d^{\dag}d +\beta(c_2^{\dag}+c_2)
 \left[(d^{\dag}+d)^2 \right.\right.\nonumber \\\left. \left.-(d^{\dag 2}d+d^{\dag}d^2 )/\sqrt j\right]/2
- \sqrt 2\beta^2\left(1+1/4j\right)\left(d^{\dag 2}d +d^{\dag}d^2
\right)/\Omega_2 \right)\, . \qquad \label{H22}
\end{eqnarray}

In Hamiltonian (\ref{H21}), the level polaron is linearly coupled
with the oscillator \nolinebreak $1$. The nonlinear selfinteracting
term $\sim d^{\dag 2}+d^2$ can be transformed out by the unitary
operator $S= \exp [r(d^{\dag 2}-d^2)]$ using the identities
\begin{eqnarray}
\tilde{d^{\dag}}\tilde d\equiv  S^{-1}  d^{\dag}d S= d^{\dag} d\cosh
4r + \sinh^2 2r + (d^2+d^{\dag 2})\frac{1}{2}\sinh 4r, \nonumber \\
(\tilde{d^{\dag}}+\tilde d) \equiv
S^{-1}(d^{\dag}+d)S=(d^{\dag}+d)e^{-2r}. \label{S}
 \end{eqnarray}
By comparing (\ref{H21}) and (\ref{S}) one obtains selfconsistently
the renormalized  frequency  $ \omega_2={\displaystyle
\frac{4\beta^2}{\Omega_2\cosh 4r}}(1+1/4j)$ and the interaction
parameter $\kappa=\sqrt 2\alpha e^{2r}$. The value of the squeezing
parameter $r$ is given by $\tanh(4r)= 2\beta^2/\Omega_2^2$. Up to
the terms of the order $j^{0}$ we get

\begin{eqnarray}
\tilde{H_{2}^{(1)}}\equiv S H_{2}^{(1)}S^{-1}= \Omega_1\left(
c_1^{\dag}c_1 +1/2\right) + \Omega_2\left( c_2^{\dag}c_2 +1/2\right)
+ \omega_2
 \tilde d^{\dag} \tilde d\nonumber\\-\kappa (c_1^{\dag}+c_1)( \tilde d^{\dag}+ \tilde
 d)-4\beta^2
 \sinh^2 2 r/\Omega_2 \,. \label{tilde}
\end{eqnarray}

As a result, the quantum fluctuations $\sim d^{\dag 2}+d^2$
renormalize the frequency $\Omega_2\rightarrow \omega_2$,
interaction $\alpha\rightarrow\kappa  $ and  the ground state
(\ref{tilde}).

 Diagonalization  of (\ref{tilde}) results in three effective independent
 oscillators $C_1, C_3$ and $c_2$, the last one remaining free,
 \begin{equation}
 H_{2d}= \epsilon_{2,
+}C_3^{\dag}C_3 +  \epsilon_{2, -}C_1^{\dag}C_1+\Omega_2
c_2^{\dag}c_2 -4\beta^2 \sinh^2 2r/\Omega_2+(\Omega_1+\Omega_2)/2\,,
\label{E2}
\end{equation}
where
\begin{equation}
\epsilon_{2, \pm}^2= \frac{1}{2}\left(\Omega_1^2+ \omega_2^2\pm
\left[(\Omega_1^2-\omega_2^2)^2+ 64 \alpha^2 \beta^2 \e^{4r}
\Omega_1/\Omega_2 \right]^{1/2}\right)\,. \label{eps2}
\end{equation}
  From (\ref{eps2}) one obtains the condition for
stability of the linear phase
\begin{equation}
\beta^2/\Omega_2 >  \alpha^2 e^{4r}/\Omega_1. \label{beta>alpha}
\end{equation}

From (\ref{beta>alpha}) we receive the  nonsymmetry  of the
radiation vs. the "normal" domain due to the squeezing: suppression
of the radiation domain. The non-symmetry of both branches of the
excitation energy  is also evident from the numerical results in
Figure \ref{boson-fig1} from the exact Hamiltonian (\ref{h}).

 The analysis of both domains shows that within linear
approximation the oscillator $c_1$ in the radiation domain plays
qualitatively the same role as does the oscillator $c_2$ in the
normal domain with simultaneous interchange of the polaron
frequencies $\omega_{1} \leftrightarrow\omega_{2}$  and coupling
constants $\alpha \sqrt{\Omega_2} \leftrightarrow
\beta\sqrt{\Omega_1}$. Hence, in linear approximation two of the
oscillators mix to two-dimensional effective oscillator while the
remaining one is decoupled. Linear analysis in adiabatic
approximation of subsections \ref{ground:1} and \ref{ground:2} yields an {\it acoustic
analogue} of the Dicke phase transition between the normal and
super-radiant phases.

\subsection{Case 3. Intermediate
domain of mixed "quasi-normal" and "radiation" domains  }
\label{ground:3}

A qualitatively new situation occurs in the nonlinear regime at
finite $j$,  when all three oscillators couple via interplay of
fluctuations due to  finite j. This situation is represented by  the
solution (\ref{sol3}).
 This solution  yields for Hamiltonian (\ref{H})
\begin{eqnarray}
 H_3 = \Omega_1\left( c_1^{\dag}c_1+1/2 \right) + \Omega_2\left(
c_2^{\dag}c_2 +1/2\right)+
\frac{4\alpha^2\bar{\mu}}{\Omega_1}d^{\dag}d
+ \frac{2\alpha^2}{\Omega_1} (1-\bar{\mu}) (d^{\dag}+d)^2 \nonumber \\
-2\alpha\sqrt{1-\bar{\mu}}(c_1^{\dag}+c_1)(d^{\dag}+d)
+\beta\frac{2\bar{\mu}-1}{\sqrt{\bar{\mu}}}
(c_2^{\dag}+c_2)(d^{\dag}+d) \nonumber \\
+ \frac{1}{\sqrt{2j}}\left[ 2\alpha(c_1^{\dag}+c_1)d^{\dag}d +\beta
\frac{\sqrt{1-\bar{\mu}}}{2\sqrt{\bar{\mu}}}(c_2^{\dag}+c_2)d^{\dag}d
+\frac{\beta}{2} \frac{\sqrt{1-\bar{\mu}}}{\sqrt{\bar{\mu}}}
 (c_2^{\dag}+c_2)(d^{\dag}+d)^2 \right.\nonumber  \\
\left.  - \frac{2\alpha^2}{\Omega_1}(1-\bar{\mu})^{1/2}(d^{\dag
 2}d+d^{\dag}d^2)-\frac{\beta}{2\sqrt{2j}\sqrt{\bar{\mu}}}(c_2^{\dag}+c_2)(d^{\dag
2}d+d^{\dag}d^2)\right]  \nonumber \\
-\frac{2\alpha^2}{\Omega_1}j(1+4\bar\mu(1-\bar\mu))+\frac{8\alpha^4
\Omega_2 \, j}{\beta^2\, \Omega_1^2}\bar{\mu}(1-\bar{\mu})
-\frac{2\alpha^2(1-\bar\mu)}{\Omega_1} \,,  \label{H3}
\end{eqnarray}
where $\bar{\mu}={\displaystyle \frac{\beta^2\,\Omega_1}{8j\Big(\alpha^2\, \Omega_2
-\beta^2\, \Omega_1 \Big)}} <1$, $\alpha > \beta
\sqrt{\displaystyle \frac{\Omega_1}{\Omega_2}}\equiv\bar\beta\,$.

Hamiltonian (\ref{H3}) implies a complex nonlinear interplay of the
quantum oscillators. The dressing of all oscillators due to the
finite $j$ is evident; for example, for the level boson $d$ its
effective frequency can be estimated by approximate quasiclassical
analysis as $\omega=
\frac{4\alpha^2}{\Omega_1}\left(1-\frac{\beta^2\Omega_1}{\alpha^2\Omega_2}\right)^{1/2}|(1-2\bar{\mu})|
$.

In the intermediate quantum domain between the quasi-normal and the
reduced radiation domain, three polarons, the dressed level
boson $d$, the dressed bosons $1$ and $2$ mix due to the finite-size
fluctuations.

\begin{SCfigure}[][htb]
\includegraphics[width=0.5\hsize]{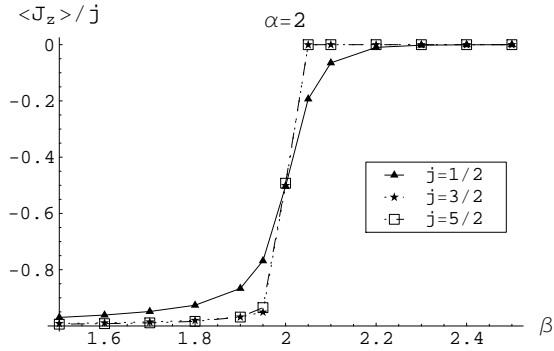}
\caption{  Order parameter $\langle J_z\rangle/j$ for $j=1/2, 3/2,
5/2$ and $\Omega_1=\Omega_2$. The finite-size effect ($j$ finite) of
the mixed domain about the crossover  vanishes asymptotically for
$j\rightarrow \infty$.}\label{boson-fig2}
 \end{SCfigure}

Numerical evaluation of the order parameter $J_z $ from the exact
Hamiltonian (\ref{h}) in Figure \ref{boson-fig2} illustrates the
existence of the intermediate phase as a partial occupation of the
 radiation domain in the neighborhood of the point
$\alpha=\bar{\beta}$ even within the normal domain
$\alpha>\bar{\beta}$. The  difference of frequencies $\Omega_1\neq
\Omega_2$ only renormalizes the position of the asymptotic
transition point in Figures \ref{boson-fig1}, \ref{boson-fig2} by
broadening the domain of the smaller frequency: the transition point
shifted inwards the domain with larger $\Omega_i$.

\section{The symmetry breaking  by semiclassical approach}
\label{semiclas}

 Linear analysis of the previous section enabled us to
 investigate each of the domains separately. As the next step,
 in (\ref{H}) we include  small nonlinear terms  up to $O(1/j)$
 to explore the finite-size effects.

Heisenberg equations related to (\ref{H1}) for the case 1 read as
follows
\begin{eqnarray}
& i \dot{c_1}= \Omega_1 c_1+\frac{2\alpha}{\sqrt {2j}}d^{\dag}d,
 \qquad \qquad  i \dot{c_2}= \Omega_2 c_2+ \beta
(d^{\dag}+d)-\frac{\beta}{4j}(d^{\dag 2}d +d^{\dag}d^2 ),
\nonumber\\
& i \dot d=\frac{4\alpha^2}{\Omega_1}d
+\frac{2\alpha}{\sqrt{2j}}(c_1^{\dag}+ c_1)d+\beta
(c_2^{\dag}+c_2)-\frac{\beta}{4j}(c_2^{\dag}+c_2)
(2d^{\dag}d+d^2)\,. \label{DE1}
\end{eqnarray}

 For strong interaction it is plausible to assume
$\omega_1={\displaystyle \frac{4\alpha^2}{\Omega_1}}\gg\Omega_1,\
\Omega_2$. Then, we can apply the adiabatic approximation neglecting
intrinsic dynamics of the level polaron $d$ of high frequency
$\omega_1$, $\dot d= 0$. Consequently, it can be eliminated, so that
it influences only implicitly the dynamics of the modes $q_1=
(2\Omega_1)^{-1/2}\langle c_1^{\dag}+ c_1\rangle $ and $ q_2=(2
\Omega_2)^{-1/2}\langle c_2^{\dag}+ c_2\rangle$. We can use the
stationary expression for $q= (8\alpha^2/\Omega_1)^{-1/2}\langle
d^{\dag}+ d\rangle$ from (\ref{DE1}). Then, the dynamic equations
for $q_1$ and $q_2$, up to the lowest order term in $1/\sqrt j$ read
\begin{eqnarray}
&\ddot q_1=  -\Omega_1^2 q_1 -\frac{\beta^2\Omega_1\sqrt
{\Omega_1}\Omega_2}{4\alpha^3 \sqrt j} q_2^2, \label{qq:1}\\
&\ddot{q_2}=  -\Omega_c^2 q_2-
\frac{\beta^2\Omega_1^2\sqrt{\Omega_1}}{\alpha^3 \sqrt{2 j}}q_1 q_2\
, \label{qq:2}
\end{eqnarray}
 where  $\Omega_c=\Omega _2\left(1-\bar{\beta}^2/\alpha^2\right)^{1/2}$, ($\alpha>\bar{\beta}$).

 We received effectively softened  frequency of the oscillator $q_2$ which justifies
 again the use of the slaving principle for the oscillator $q_1$,
$\Omega_1^2\gg\Omega_1^2(1-\bar{\beta}^2/\alpha^2)$. Its dynamics is
then implicitly ordered by the dynamics of the oscillator $q_1$.
Then, similarly as in the previous case, we obtain
\begin{equation}
\ddot{q_2}= -\frac{d V(q_2)}{d q_2}\,, \label{dV}
\end{equation}
where
\begin{equation}
V(q_2)=\frac{1}{2}\Omega_c^2 q_2^2
+\frac{\beta^5\Omega_1^2\Omega_2}{16\alpha^6 j\sqrt{2}}q_2^4+V_0\,.
\label{V}
\end{equation}
Thus, for finite $j$, equation (\ref{dV}) for the fluctuations $q_2$
in the normal domain represents the ideal Duffing oscillator. The
potential (\ref{V}) implies a stable "normal" phase up to the
condition $\alpha=\bar{\beta}\equiv \beta\sqrt{\Omega_1/\Omega_2} $
to the order $1/j$ for each $j$.

We conclude that in the adiabatic approach in the normal domain
 there occur finite-size fluctuations and instability of the
oscillator (\ref{qq:2}) because of softening its frequency
$\bar{\omega_2}$; the softening here represents approaching closer
 the radiation domain.

Let us suppose that the energies at both sides of the transition
between the normal and the radiation domains are symmetric
 when interchanging $\alpha\leftrightarrow \beta $
and $\Omega_1\leftrightarrow \Omega_2$, ($r\rightarrow 0$). In the
radiation domain the nonlinearity yields one-instanton solution
\begin{equation}
\bar{q_{21}} (\tau-\tau_0)=
\pm\frac{2\alpha_1^3}{\bar{\beta}^2\Omega_1}
\left[\frac{2j}{\Omega_1}\left(
1-\alpha^2/\bar{\beta}^2\right)\right]^{1/2}\tanh\left(\frac{\Omega_1}{\sqrt
2}\left
(1-\alpha^2/\bar{\beta}^2\right)^{1/2}(\tau-\tau_0)\right)\,.
\label{K1}
\end{equation}

Here, $\bar {q_2}= q_2\sqrt {\frac{\Omega_1}{\Omega_2}}, \ \bar
{\beta}= \beta\sqrt {\frac{\Omega_1}{\Omega_2}}$.  The one-instanton
solution (\ref{K1}) is associated with the tunneling between the
extrema of the potential inverted to (\ref{V}) if $\bar{\beta}>
\alpha$ and $\tau\rightarrow it$. Hence, at finite $j'$s  there
appears new instanton phase at the maximum softening of the
frequency at $\bar{\beta} \rightarrow \alpha$ in the radiation
domain. Generally, there occurs a sequence of repeated tunnelings
(oscillations between two equivalent minima of a local potential)
for each lattice site.
 Moreover,  there exists the coupling between the oscillators $1$
and $2$ (\ref{qq:1},\ \ref{qq:2}) which was neglected in the linear
 approximation (\ref{dV}). In fact, within more subtle
calculations there would occur tunnelings mediated by two coupled
oscillators (one of them being a polaron) for each $j$ instead of
one of the adiabatic treatment.

\section{Statistical characteristics for excited boson complex spectra} \label{stat}

 For $\alpha=0$ or $\beta=0$, the excited boson spectrum of the Hamiltonian (\ref{h}) is the
superposition of two sets of levels of Fock harmonic oscillators. If
$\alpha\neq\beta\neq 0$, respective wave functions of
 excited states become to overlap and the nearest levels strongly repulse
 which results in multiple level-avoidings. The spectrum acquires
 respective degree of complexity, i.e., of quantum chaotic behaviour.
 For the excited complex spectra of atoms, molecules and
nuclei \cite{Guhr,Dyson},  several spectral characteristics as
signatures of quantum chaos were introduced.
  The mostly used one is the probability distribution function $P(S)$ of the
nearest-neighbour level spacings (NNLS) utilized as a standard
testing point for investigation of the issues in spectral quantum
chaology \cite{Guhr}. The excited spectra of a supposedly quantum
chaotic system can be modeled in terms of statistical ensembles for
which the random matrix theory specifies several classes of
universal probability distributions (Wigner surmise) of the NNLS
dependent solely on the symmetry of the underlying Hamiltonians
\cite{Dyson}. In the case of the Dicke model the  NNLS distribution
is namely the Wigner-Dyson one $P(S)\sim S \exp(- S^2)$ which is the
sign of fully developed quantum chaos
\cite{Emary:2003:b,Dyson}. In the absence of overlapping of the wave
functions, respective levels become to cross each other and the
Poisson distributions $P(S)\sim\exp(-S)$ result.

Recently we have reported the results on NNLS distribution for
excited boson spectrum of the JT model (taken as function of coupling constants
$\alpha$, $\beta$)
\cite{Majernikova:2008,Majernikova:2006:b} which is the case
$j=1/2$, $\Omega_1=\Omega_2$, $\alpha\neq\beta$ in the notation of
(\ref{h1}) of the present paper. These distributions essentially
deviate from the prediction of the Wigner-Dyson form.  The
peculiarity of the JT system of a single molecule with
equal frequencies was that the Wigner-Dyson distribution for $P(S)$
has never been reached, but the distribution close to the
semi-Poisson form $P(S)= 4S \exp (-2S)$ appeared as the most typical
``chaotic'' one.

In this section we present statistical characteristics (NNLS
distributions, entropy of level occupation and spectral density of
states) for $\Omega_1\neq\Omega_2$, $j\geq 1/2$. We solved
numerically the eigenvalue problem for quantum Hamiltonian
({\ref{h}). This Hamiltonian was diagonalized with the basis of the
boson Fock states for bosons $1$ and $2$. An inevitable truncation
error was produced when taking only $N_1$ and $N_2$ {\it boson} Fock
states for each of $2j+1$ {\it electron} levels, so that the results
were checked against the convergence (with changing the numbers
$N_1, N_2$). Only about $\sim 1100$ lower states were used for
calculation of the statistics out of typically $\sim (8\div 10)
\times 10^3$. The raw energy spectrum obtained had to be treated by
an unfolding procedure in order to ensure the homogeneity of the
spectrum (constant local density of levels). Thereafter the
statistical data were gathered in a standard fashion from small
intervals in the space of parameters $(\alpha, \beta)$. The
calculations were performed essentially along the same lines as our
previous calculations for the Jahn-Teller problem with $j=1/2$
\cite{Majernikova:2008,Majernikova:2006:b} where additional details
are given as to the convergence check, unfolding procedure and
gathering statistics.

  The results for the level spacing distributions for different phonon
frequencies $\Omega_1\neq \Omega_2$ and $j>1/2$ show rather vast
variety ranging from the Wigner-Dyson to the Poisson distribution as
limiting cases but recovering also the semi-Poisson distribution in
a rather wide range of model parameters. With increasing $j$,
general tendency consists in increasing the range of parameters
where the NNLS shows deviations from the Poisson form towards
chaoticity up to its maximum degree of the Wigner-Dyson form. For
moderate $j$ ($j=7/2$ and $\Omega_1\neq \Omega_2$ exemplified in
Figure \ref{boson-fig3}) there appears a well-marked domain of the
Wigner-Dyson distribution. For example, for $\Omega_2/\Omega_1=2$
this domain stretches for the values of parameters $1< \alpha,\beta<
3$.

\begin{figure}[htb]
\includegraphics[width=0.8\hsize]{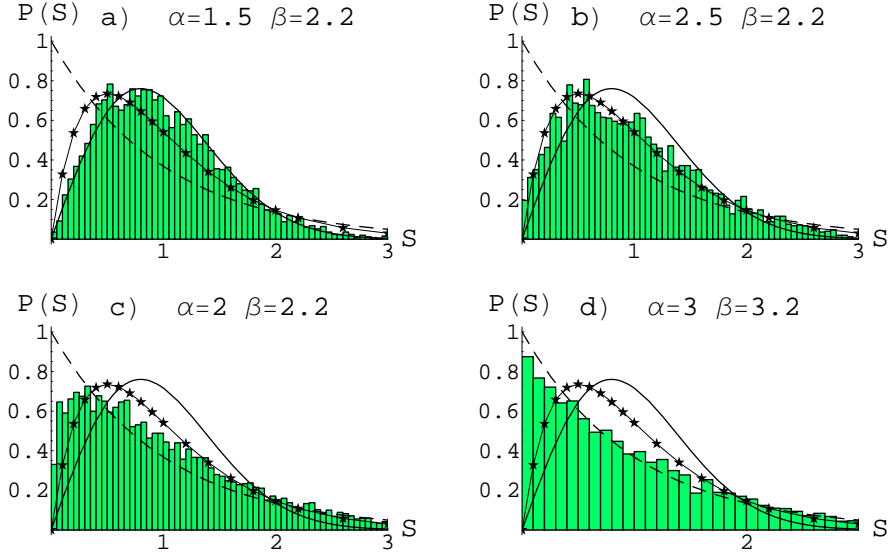}
\caption{ Level spacing distributions for $j=7/2$ and different
($\alpha$, $\beta$) (parameters are scaled to $\Omega_1=1$). Upper
row ((a,b)): $\Omega_2= 2\,\Omega_1$; bottom row ((c,d)): $\Omega_2=
0.5\,\Omega_1$. The curves show Wigner-Dyson (solid), Poisson
(dashed) and semi-Poisson (stars) distributions.
 NNLS distributions in b), c) are close to the semi-Poisson distribution $P(S)=4 S \exp (-2S)$
\cite{Majernikova:2006:b}; histograms in a) and d) are almost
perfect Wigner-Dyson ($P(S)\sim S\exp(-S^2)$ and Poisson $P(S)
\sim\exp (-S)$ distributions. The distributions are $j-$dependent,
that is an evidence of the interference of both kinds of
fluctuations: finite-size ones and those of the excited level
spacings} \label{boson-fig3}
\end{figure}

Therefore the impact of the second boson mode is that it essentially
changes the quantum statistics. For the non-resonance case
$\Omega_1\neq\Omega_2$ (Figure \ref{boson-fig3}) we observe a
considerable suppression of chaos manifesting in reduction of the
``pure'' domain of the Wigner-Dyson chaos in the space of the
parameters $\alpha,\beta$. In the resonance case $\Omega_1=\Omega_2$
basically similar results as in our previous analysis
 \cite{Majernikova:2006:b} are recovered, the most important one
being the complete disappearance of the domain of the Wigner-Dyson
NNLS distribution. According to the results of Section \ref{ground} the
reduction of the Wigner-Dyson chaos is to ascribe to quantum
correlations (squeezing) in the radiation domain
and the correlation (entanglement) of both the bosons.

Present results essentially deviate from the Wigner-Dyson statistics
imposed by one-boson Dicke model \cite{Emary:2003:b}. For
comparison, we performed the same calculations of NNLS distributions
for the standard  Dicke model (Figure \ref{boson-fig4}). The
distributions follow closely the Wigner-Dyson form of the $P(S)$
curve independent of the choice of  pairs of parameters $\omega_0$
and $\lambda$ in both the subcritical (normal) and supercritical
(radiation) regions.  Good agreement is achieved only for high
values of $j$, finite-size fluctuations occur for small $j$.

\begin{figure}[htb]
\includegraphics[width=0.7\hsize] {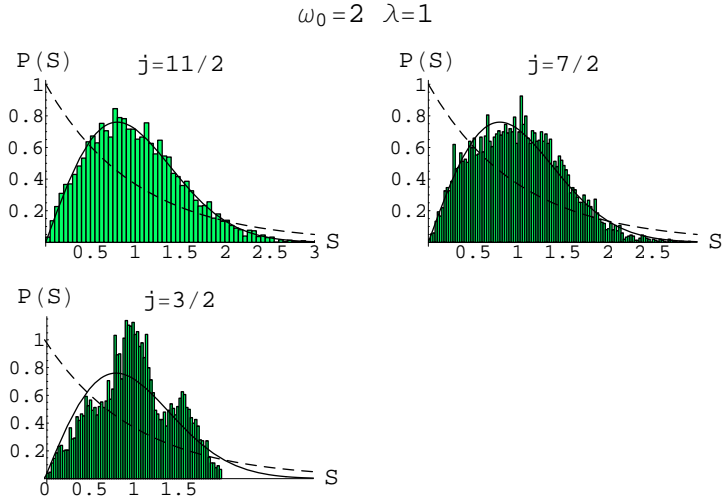}
\caption{ Level spacing distributions for one-boson Dicke model
(\ref{HD})  (boson frequency $\Omega$ scaled to $1$). The parameters
of the Dicke model (\ref{HD}) correspond to our parameters $\alpha
$, $\beta$ as $\lambda\rightarrow \sqrt {\Omega_1/\Omega_2}\beta$,
$\omega_0\rightarrow 4\alpha^2/\Omega_1$. For sufficiently large
$j$, the  distribution is approaching the Wigner-Dyson distribution
for any pairs of the parameters $\lambda$, $\omega_0$ as one can see
for $j=7/2, 11/2$. The smaller $j$, the stronger are the
nonuniversal fluctuations. Similarly to the previous Figure the
j-dependence is the evidence for mixing of both kinds of
fluctuations in the Dicke model. } \label{boson-fig4}
\end{figure}

 To visualize the wave functions we  considered their spreading over the
electron levels and integrated out two boson degrees of freedom.
Thus we found numerically the probability $P_i^{(n)}$ of the wave
function $\phi_i^{(n)}$ to occupy a given electronic level $i$ of
the system. The full wavefunction depending on both vibron and
electron level variables is calculated by the same scheme of the
diagonalization of the quantum Hamiltonian in the representation of
Fock states as described above. If $\beta=0$, Hamiltonian equation
is split onto $N(=2j+1)$ independent equations labeled by
$i=1,\dots,(2j+1)$, so that the wavefunction for each given state
$n$ is localized on some electron level $i$. If $\beta\neq 0$,
Hamiltonian matrix is no more diagonal (since $J_x$ is nondiagonal),
so that its components get interacting. The resulting eigenfunction
is in general $(2j+1)$-fold function occupying each electronic level
(see Section \ref{model} for the discussion on normal and super-radiant phase
in the thermodynamic limit). Let $\chi_i^{(n)}(Q_1,Q_2)$ be the i-th
component of the $(2j+1)$-dimensional vector of the eigensolution of
Hamiltonian matrix equation for n-th energy level in the
"coordinate" representation ($\hat{Q}_l\propto b_l^{\dag} +b_l$,
$l=1,2$). Then the
  occupation probabilities  of the i-th electronic level are
  $P_i^{(n)}=\int\int
|\chi_i^{(n)}|^2 \mathrm{d} Q_1 \mathrm{d} Q_2$. In this
representation  we exemplify the wave functions in Figure
\ref{boson-fig5} for the levels $n=1,4$ and the parameters around
the point of QPT $\alpha=2$,  $\beta=1.95, 2.05$ and
$\Omega_1=\Omega_2$. The abrupt change in the shape of wavefunctions
when going from normal to super-radiant phase is easily perceivable
not only for the ground state, but for lowest excited states as well
(in this example for n=4). However, for higher excited states the
wavefunctions generally spread over the whole available electron
space, irrespective to the relation between $\alpha$ and $\beta$.
The last relation  is also reflected in the statistical properties
of levels: they are invariant with respect to the  exchange $\alpha
\leftrightarrow \beta$, $\Omega_1\leftrightarrow \Omega_2$, likewise
it was already shown for the generalized JT model
\cite{Majernikova:2006:b}.

\begin{SCfigure}[50][htb]
\includegraphics[width=0.6\hsize]{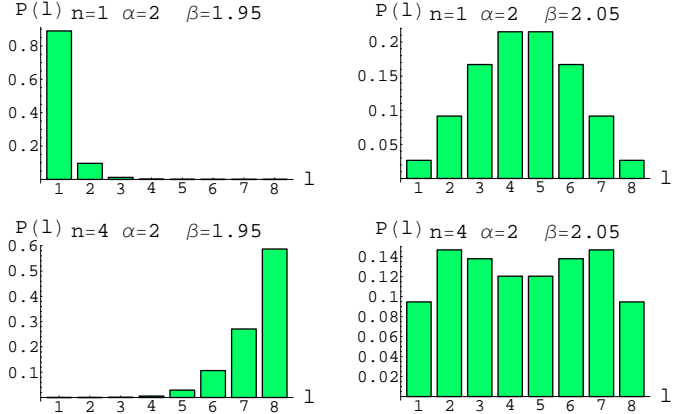} \hspace{-5em}
\caption{The occupations $P(l)$ of the electronic levels $l= 1,\dots
2j+1$ (see Sect. V) close the the critical point $\alpha=\beta$
($\Omega_1=\Omega_2=1$; $j=7/2$) of the crossover between the
quasi-normal and radiation domains for the ground $n=1$ and fourth
$n=4$ excited state. For $\Omega_1\neq\Omega_2$ the occupations
remain qualitatively the same. } \label{boson-fig5}
\end{SCfigure}

In order to characterize quantitatively the
 behavior of the wavefunctions of excited states we introduce the
entropy of level occupation
\begin{equation}
S_n= - \sum \limits_{i=1}^N P_i^{(n)} \log P_i^{(n)} \,,
 \label{Sn}
\end{equation}
where $\sum\limits_{i=1}^N P_i^{(n)} =1$. In Figure \ref{boson-fig6}
for $\Omega_1=\Omega_2 $ we plot the said entropies for first 1000
quantum levels of a system with four electron levels ($j=3/2$). One
can see from these figures that in the most excited states  the
electronic levels are equally populated (with probabilities
$1/(2j+1)$, e.g., last item in Figure (\ref{boson-fig5})),  thus the
said entropies tend to their limiting value $\log (2j+1)$. For four
levels the entropy yields the value $S_4 = 2\log 2= 1.38$ which is
confirmed by Figure \ref{boson-fig6} for $\alpha\neq \beta$.
However, among these extended states there emerge ``localized''
levels of lower values of the entropy which are characterized by
relative localization of wavefunction on smaller number of electron
levels. Such levels form marked branches seen in Figure
\ref{boson-fig6}, which remind the branches of ``exotic''
(localized) states similar to those found in our preceding paper
\cite{Majernikova:2006:a}. Most of the states show an intermediate
degree of localization between both the limits. These intermediate
electronic states correspond to the critical mixed
domain of the normal and radiation ones.

\begin{figure}[htb]
\includegraphics[width=0.9\hsize]{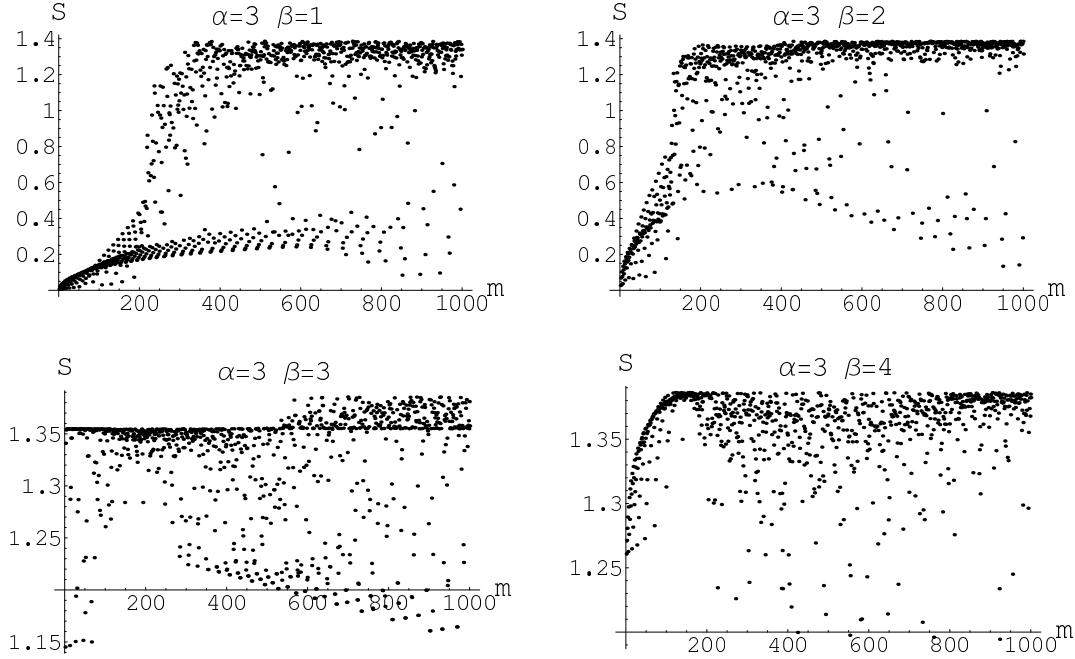}
\caption{Entropies of occupation of electronic levels (\ref{Sn}) as
function of the number of the excited state $m$ for $j=3/2$ ($4$
levels). The states with entropy lower than the limiting value $\log
(2j+1)$ have larger measure of localization.   } \label{boson-fig6}
\end{figure}

Another way to  visualize the complex structure of the excited
states and corresponding wave functions is the representation of the
spectral density given by the imaginary part of the projected
resolvent \cite{Ziegler:2005}.  It shows the characteristic
frequencies of the final state of the evolution of a system starting
from the projected Hilbert space and returning to it. The spectral
density of states $F(E)$ can be defined with respect to some
initially prepared state of the wave packet $|\Psi_0\rangle$  and is
related to the return probability to this state in the course of the
system evolution through the exact states $\Phi_n$, eigenfunctions
of the energies $E_n$ of the system at small $\varepsilon \simeq 0$:

\begin{equation}
F(E)\equiv \mathrm{Im}
 \langle \Psi_0 |(E-\hat{H}-i\varepsilon)^{-1}
|\Psi_0\rangle = \varepsilon\sum_n \frac{|\langle
\Phi_n|\Psi_0\rangle |^2}{(E-E_n)^2+\varepsilon^2}, \label{spectfun}
\end{equation}
where $|\Phi_n\rangle$ and $E_n$ are the eigenfunctions and
corresponding eigenvalues of Hamiltonian $\hat{H}$. The small
parameter $\varepsilon$ fixes the rules of handling the poles of the
Green function in the complex space. In Figure \ref{boson-fig7} we
show the examples of the spectral density for the values of
parameters $\alpha$, $\beta$ below and above the critical line. As
the initial state $|\Psi_0\rangle$ we took the ground state with
$\alpha=\beta=0$, that is the product of the electronic Dicke state
$|j,-j\rangle_{el}$ and the phonon state $|0,0\rangle_{ph}$ with
zero number of both bosons 1 and 2. Thus the spectral density in
Figure (\ref{boson-fig7}) characterizes the evolution of the system
with the interactions $\alpha$, $\beta$ switched on in the initial
time. The peaks of each $E_n$ measure overlap between the initial
state $\Psi_0$ and the eigenstate $\Phi_n$ for $\varepsilon\sim  0$.

\begin{SCfigure}[][htb]
\includegraphics[width=0.5\hsize]{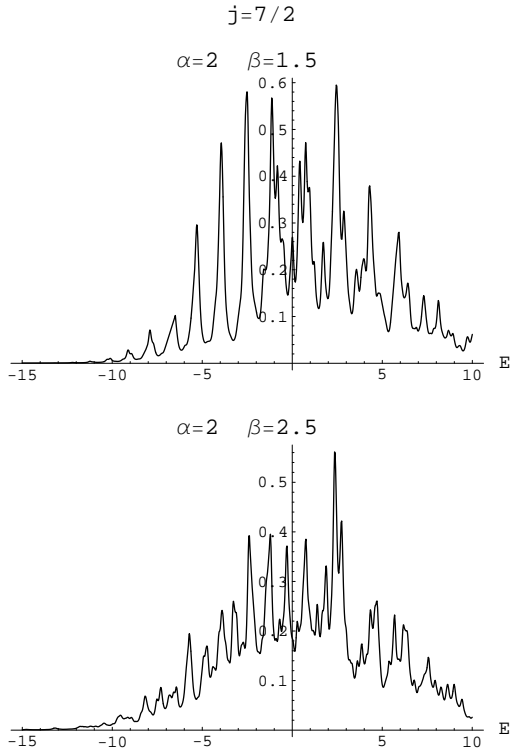}
\caption{Examples of the spectral density of states $F(E)$
(\ref{spectfun}) for $\alpha=2$ and $\beta=1.5, 2.5$ (We plot the
resonant case $\Omega_1=\Omega_2=1$; $j=7/2$; the non-resonant cases
do not change the picture qualitatively. The same applies for
previous two figures.)  } \label{boson-fig7}
\end{SCfigure}

\section{Conclusion}
\label{concl}

 We performed analytical and numerical analyses of the multispin JT model
 with fully broken local rotation symmetry ($\alpha\neq\beta$ and
$\Omega_1\neq\Omega_2$).

Here, we will briefly summarize relations between the effects and
their causes especially on the phase transition and statistical
characteristics:

(i) Fluctuations in electron subsystem - finite-size fluctuations
for small $j$ occur mainly in the
 region close to the normal-radiation phase
transition, Figs.~\ref{boson-fig1},~\ref{boson-fig2}. Occupation of
 electronic levels close to the transition exhibits a partial
excitation of the normal and dephasing of the radiation phase,
 Fig.~\ref{boson-fig5}.
Thus, they smooth the phase transition point to a crossover region
between the normal and radiation domains as a rather broad
intermediate critical "mixed" domain. Similar finite-size effect for
small  $j$ exhibits also the order parameter in the electron space
$\langle J_z\rangle/j$,  Fig.\ref{boson-fig2}.

The crossover of the wave functions: the excited states
  are spread over small number of sites in the
"normal" domain to the macroscopic distribution of the excitation
over the lattice in the radiation domain, Fig.\ref{boson-fig5}.

(ii) Fluctuations in the {\it excited boson subsystem}: The
fluctuations of the nearest neighbour levels spacings
 are characterized by the distributions $P(s)$.
Numerical results for $P(s)$ (Fig.\ref{boson-fig5}) show markable
perturbations for small values of the finite-size lattice parameter
$j$. This is an evidence for the interplay of both kinds of
fluctuations. In a semiclassical limit $j\rightarrow \infty $ the
perturbations of $P(s)$ due to $j$ vanish and the distributions are
determined solely by the level repulsions (avoidings) moderated by
the level correlations.

(iii) The difference in mode frequencies essentially contributes
(besides the difference of the coupling constants) to the departure
from the rotational symmetry  of the underlying Hamiltonian. The
additional (antisymmetric) boson mode mediates  {\it correlations,
mode squeezing and entanglement}  by an assistance of the tunneling
between the levels that weaken the level repulsions at small level spacings. They cause the
nonuniversality of the distributions $P(s)$ which is absent  in the
one-boson Dicke model.

The Holstein-Primakoff (HP) bosonization Ansatz in Section \ref{ground} makes it
possible to describe the system  as a system of three interacting
bosons and find some useful, though approximate, analytical results.
  For a finite $j$, there opens a critical quantum "mixed" domain
  with partial occupation of the excitation space of all
three coupled oscillators. This domain exists in the discrete Dicke
model as well and vanishes in the semiclassical limit
$j\rightarrow\infty$.

The effective oscillators enable us to extract a sequence of repeated
tunnelings in the radiation domain. The
 domain can be thus represented as an almost ideal
instanton--anti-instanton phase (instanton lattice). {\it Each
instanton represents a tunneling event, so that the radiation domain
is  the most chaotic one}. The dephasing mechanism in the
radiation phase here is shown being caused by additional
correlations from the neighbour instantons which perturb the ideal
periodicity of the instanton lattice.

Several applications of the present two-boson finite-lattice system
can be proposed:

(1) A zero temperature mechanism for interpretation of the coherence
dephasing and the broadening of spectra
 \cite{Kuusmann:1975,Kmiecik:1987,Kishigami:1992,Ding:1997} and the
broadening of the zero-phonon lines \cite{Malcuit:1987}.

(2) A similar impact on the properties   of ultracold atoms in
optical lattices  \cite{Ziegler:2005,Roth:2004,Fisher:1989} due to
the quantum localization-delocalization transition from the Mott
insulator to the superfluid state.
 There  the  phase coherence of Bose atoms is destroyed by the suppression of the tunneling
 due to the localization by the strong lattice potential.
 From the competition of the tunneling and the localization due to the lattice
 there results the phase transition between the highly coherent superfluid phase of Bose atoms and
  the localized Mott insulating phase of the atoms on the lattice
  \cite{Fisher:1989}. If quantum fluctuations can mediate the tunneling
  within the insulating phase, as well as perturb  coherence of
  the superfluid phase, the system of Bose atoms in optical lattices would exhibit analogous properties
  as those of the two-boson Dicke model.

(3) Assistance  of the additional coherent boson mode in the level
splitting might be utilized as a possible mean to control chaos in
related spectral properties of optical systems based on
(pseudo)spin-boson models.\\

 \ack  The
financial support by the project No. 2/0095/09 of the Grant Agency
VEGA of the Slovak Academy of Sciences is highly acknowledged.\\


\begin{thebibliography}{99}
\bibitem{Graham:1984:a}  Graham P and  H\"{o}hnerbach M 1984 {\it Z. Phys. }B {\bf 57} 233
\bibitem{Graham:1984:b}  Graham P and   H\"{o}hnerbach M {\it Phys. Lett.} A {\bf 101} 61
(1984)
\bibitem{Lew:1991} Lewenkopf  C H,   Nemes MC, Marvulle V,
Pato M P and  Wreszinski W F 1991 {\it Phys. Lett.} A {\bf 155} 113
\bibitem{Cibils:1995}  Cibils M, Cuche Y and  M\"uller G 1995 {\it Z. Phys. B} {\bf 97}
565
\bibitem{Dicke}  Dicke R H  1954 {\it Phys. Rev.} {\bf 93} 99
\bibitem{Wang:1973}  Wang Y K and  Hioe F T 1973 {\it Phys. Rev.} A {\bf
7} 831 ; Hepp K and  Lieb E H 1973 {\it Phys. Rev.} A {\bf 8}  2517
\bibitem{Littlewood:2000} Eastham P R and Littlewood P B 2000 {\it Solid State Comm.} {\bf 116} 357
\bibitem{Emary:2003:a}  Emary C and Brandes  T  2003 {\it Phys. Rev. Lett.} {\bf 90} 044101
\bibitem{Emary:2003:b} Emary C and Brandes T
 2003 {\it Phys. Rev.} E{\bf 67} 066203
 \bibitem{Haake}  Harms K-D and  Haake F 1990 {\it Z. Phys.} B {\bf 79} 159 ;
 Gnutzmann S,  Haake F and  Kus M 2000 {\it J. Phys.} A {\bf 33} 143
\bibitem{Tolkunov:2007}  Tolkunov D and Solenov D  2007 {\it Phys. Rev.} B {\bf 75} 024402
\bibitem{Pfeifer}  Pfeifer P 1982 {\it Phys. Rev.} A {\bf 26} 701
\bibitem{Larson:2008}  Larson J 2008 {\it Phys. Rev. } A {\bf 78} 033833
\bibitem{Guhr} Guhr T, M\"uller-Groeling A, Weidenm\"uller H A 1998 {\it Phys. Rep.} {\bf 299} 190
\bibitem{Kasprzak:2006}  Kasprzak J, Richard M, Kundermann S, Baas A, Jeambrun P, Keeling J M J,
Marchetti F M, Szyma M H, Andr\'e R, Staehli J L, Savona V,
Littlewood P B, Deveaud B and  Dang Le Si  2006 {\it Nature} {\bf
443} 409
\bibitem{Eastham:2006} Eastham P R and Littlewood P B 2006 {\it Phys. Rev. B} {\bf 73} 085306
\bibitem{Keeling:2007} Keeling J 2007 {\it J. Phys.: Cond. Matter }
{\bf 19} 295213
\bibitem{Love:2008} Love A P D, Krizhanowskii D N, Whittaker D M,
Bouchekioua R, Sanvitto D, Al Rizeigi S, Bradley R, Skolnick M S,
Eastham P R, Andr\'e R and Dang Le Si 2008 {\it Phys. Rev. Lett.}
{\bf 101} 067404
\bibitem{Dyson} Dyson F J and Mehta M L 1963 {\it J. Math. Phys. }{\bf 4
} 701
\bibitem{Majernikova:2002}  Majern\'{\i}kov\'a E and  Shpyrko S
2002, {\it Phys. Rev. B }{\bf 65} 174305
\bibitem{Majernikova:2003}  Majern\'{\i}kov\'a E and  Shpyrko S 2003
{\it J. Phys.: Cond. Matter} {\bf 15} 2137
\bibitem{Majernikova:2008}  Majern\'{\i}kov\'a E and Shpyrko  S 2008
 {\it J. Phys. A: Math. Theor.} {\bf 41} 155102
 \bibitem{Majernikova:2006:a}  Majern\'{\i}kov\'a E and   Shpyrko S 2006
{\it Phys. Rev.} E {\bf 73} 066215
\bibitem{Ziegler:2005} Ziegler K 2005 {\it Phys. Rev.} B {\bf 72} 075120
\bibitem{Roth:2004} R. Roth and K. Barnett 2004 {\it J. Phys. B:
At. Mol. Opt. Phys. } {\bf 37} 3893
 \bibitem{Vidal:2006} Vidal J and  Dusuel S 2006 {\it Europhys. Lett.} {\bf 74} 817
 \bibitem{Yuen} Yuen H P 1976 {\it Phys. Rev. A } {\bf 13 } 2226;
Loudon R and Knight P L 1987 {\it J. Mod. Opt.} {\bf 34} 709-754
\bibitem{Lowen:1988} Gerlach B and  L\"owen H 1988 {\it  Phys. Rev. B }{\bf 37} 8042 ;
1991  {\it Rev. Mod. Phys.} {\bf 63} 63
\bibitem{Eisert:2004} Eisert J,  Plenio M B,  Bose S and
Hartley J 2004 {\it Phys. Rev. Lett.} {\bf 93} 190402
\bibitem{Long:1958}  Longuet-Higgins H C,  \"Opik U, and  Pryce M H L 1958 {\it Proc.
 Roy. Soc. London, Ser.} {\bf A244} 1
\bibitem{Kuusmann:1975}  Kuusmann I L,  Liblik P K  and
Lushchik Ch B 1975 {\it  JETP Lett.}{\bf 21} 72
\bibitem{Kmiecik:1987}  Kmiecik H J,  Schreiber M,  Kloiber T,
Kruse M and  Zimmerer G 1987 {\it  J. Lumin.} {\bf 38} 93
\bibitem{Kishigami:1992} Kishigami-Tsujibayashi T, Toyoda K and
T. Hayashi 1992  {\it Phys. Rev.} B {\bf 45} 13 737
\bibitem{Ding:1997} Xiaoya Ding and  Wright J C 1997 {\it Chem. Phys.
Lett.} {\bf 269} 341
\bibitem{Malcuit:1987}  Malcuit M S, Maki J J,  Simkin D J and
 Boyd R W, {\it Phys. Rev. Lett.} 1987 {\bf 59} 1189
\bibitem{Primakoff}  Holstein T and Primakoff  H 1949 {\it Phys. Rev. } {\bf 58} 1098
\bibitem{Brandes:2005}  Brandes T 2005 {\it Phys. Repts} {\bf 408} 315
 \bibitem{Majernikova:2006:b} Majern\'{\i}kov\'a E and   Shpyrko S 2006
{\it Phys. Rev.} E {\bf 73} 057202
 \bibitem{Fisher:1989}  Fisher M P A, Weichman P B,  Grinstein G and  Fisher D S 1989 {\it  Phys. Rev.}
 B {\bf 40} 546
\end{thebibliography}
\end{document}